\documentclass[twocolumn,prb,showpacs,graphicx,superscriptaddress]{revtex4}
\usepackage{graphicx}
\usepackage{dcolumn}
\usepackage{bm}
\usepackage[cmex10]{amsmath}
\usepackage[utf8]{inputenc}
\usepackage{float}
\usepackage{amssymb}
\usepackage[T1]{fontenc}
\usepackage{natbib,hyperref}

\usepackage{epsfig}

\begin{document}

\preprint{APS/123-QED}

\title{Field- and temperature-modulated spin-diode effect in a GMR nanowire with dipolar coupling}

\author{Piotr Ogrodnik}
\email{piotrogr@if.pw.edu.pl}
\affiliation{University of Michigan, Department of Electrical Engineering and Computer Science, Ann Arbor, MI 48109, USA}
\altaffiliation[permanent address: ]{Warsaw University of Technology, Faculty of Physics, , ul. Koszykowa 75, 00-662 Warszawa, Poland}

\author{Tomasz Stobiecki}
\affiliation{AGH University of Science and Technology, Department of Electronics, Al. Mickiewicza 30, 30-059 Krak\'{o}w, Poland}
\affiliation{AGH University of Science and Technology, Faculty of Physics and Applied Computer Science, Al. Mickiewicza 30, 30-059 Krak\'{o}w, Poland}

\author{J\'{o}zef Barna\'{s}}
\affiliation{Adam Mickiewicz University, Faculty of Physics, ul. Umultowska 85, 61-614 Poznań, Poland}
\affiliation{Institute of Molecular Physics, Polish Academy of Sciences, ul. M. Smoluchowskiego 17, 60-179 Poznań, Poland}

\author{Marek Frankowski}
\affiliation{AGH University of Science and Technology, Department of Electronics, Al. Mickiewicza 30, 30-059 Krak\'{o}w, Poland}

\author{Jakub Chęci\'{n}ski}
\affiliation{AGH University of Science and Technology, Department of Electronics, Al. Mickiewicza 30, 30-059 Krak\'{o}w, Poland}
\affiliation{AGH University of Science and Technology, Faculty of Physics and Applied Computer Science, Al. Mickiewicza 30, 30-059 Krak\'{o}w, Poland}

\author{Francesco Antonio Vetr\`{o}}
\affiliation{\'{E}cole Polytechnique F\'{e}d\'{e}rale de Lausanne - EPFL, Institute of Physics, Station 3 , CH-1015 Lausanne, Switzerland}

\author{Jean-Philippe Ansermet}
\affiliation{\'{E}cole Polytechnique F\'{e}d\'{e}rale de Lausanne - EPFL, Institute of Physics, Station 3 , CH-1015 Lausanne, Switzerland}

\date{\today}

\begin {abstract}
An analytical model  of the spin-diode effect induced by resonant spin-transfer torque in a ferromagnetic bilayer with strong dipolar coupling provides the resonance frequencies and the  lineshapes  of the magnetic field spectra obtained under field or laser-light modulation. The effect of laser irradiation is accounted for by introducing the temperature dependence of the saturation magnetization and anisotropy, as well as thermal spin-transfer torques. The predictions of the model are compared with experimental data obtained with single Co/Cu/Co spin valves, embedded in nanowires and produced by electrodeposition. Temperature modulation provides excellent signal-to-noise ratio. High temperature-modulation frequency is possible because these nanostructures have a very small heat capacity and are only weakly heat-sunk. The two forms of modulation give rise to qualitative differences in the spectra that are accounted for by the model.
\end{abstract}

\maketitle

\section{Introduction}
The spin diode effect (SDE) is a well established method of electrical detection of magnetic dynamics in ferromagnetic layers.\cite{harder2016electrical,tulapurkar2005spin,sankey2006spin,sankey2007measurement,kubota2007quantitative,locatelli2014spin,ziketek2015rectification,jenkins2016spin} The effect occurs when an alternating current ({\it AC}) passes through a magnetic structure and excites oscillations of the magnetization vector. This, in turn, leads to variation of the resistance due to the magnetoresistance effect. The oscillating resistance can then mix with the {\it AC} current to produce a direct ({\it DC}) output voltage, $V_{DC}$. The SDE is of great importance  from the point of view of further development of magnetic sensors, microwave communication and ultrafast electronics -- especially since the nanostructures manufactured nowadays make it possible to get smaller and more efficient electronics elements.~\cite{locatelli2014spin,ziketek2015influence,jenkins2016spin}

Until now, the SDE was investigated mainly in systems with one ferromagnetic free layer. Although more ferromagnetic layers were present in a device, they were usually assumed to be magnetically stiff (pinned layers). However, in some of the experimentally investigated structures there are two or more layers which are magnetically free and, in addition, are dynamically coupled by  the RKKY-like exchange interaction and/or by dipolar interactions.~\cite{layadiPRB,layadi2015theoretical,layadi1990ferromagnetic,ziketek2015influence,koziol2017interlayer}
Moreover, they are also dynamically coupled by spin transfer torque (STT) effects.~\cite{morijama} Nevertheless, most of these devices with multiple free magnetic layers, still include at least one, usually thick, pinned magnetic polarizer.\cite{natarajan2016high, bell2017dual}

Regardless of its source, the coupling between layers may lead, in general, to more complex magnetization dynamics and a non-trivial behaviour of the device. Therefore, it is important to have a theoretical description in order to understand the experimentally measured characteristics of these devices. In this paper, we address this problem and  focus on two mechanisms of dynamic coupling: the spin transfer torque effect and the dipolar interactions between two magnetically free layers in a giant magnetoresistance (GMR) nanowire system.

We propose an analytical model to study the SDE in such a system, which is based on magnetization dynamics described in terms of the Landau-Lifshitz-Gilbert equation. The relevance of the model is demonstrated by showing that it predicts the qualitative differences observed in the ferromagnetic resonance spectra measured under field~\cite{gonccalves2013spin} or temperature modulation.~\cite{gravier2004thermopower} We analyse in detail the influence of the type of SDE experimental modulation technique on the shape of the resonance spectra. As we consider the laser modulation technique, we assume a temperature dependence of the saturation magnetization and anisotropy parameters. Also we consider the presence of thermal STTs due to temperature gradient generated along the nanowire.

The Co/Cu/Co nanowires were fabricated by electrochemical deposition technique, and in experiments we used structures of diameters about 30 nm. In such structures there is a strong dipolar coupling between the Co layers. To support our analytical calcualtions based on the macrospin model, we also performed micromagnetic simulations.

The paper is organized as follows. In section \ref{sec:theory} we present a theoretical description of the magnetization dynamics in the system and also describe briefly the spin diode effect. In section \ref{sec:experiment} the experimental part is presented. In section \ref{sec:results} we discuss  results for the $V_{DC}$ signal and line shapes in the case of field and laser modulation techniques. Summary and final conclusions are given in section \ref{sec:concl}.

\section{Theoretical description\label{sec:theory}}

\begin{figure}
\includegraphics[width=6.5cm]{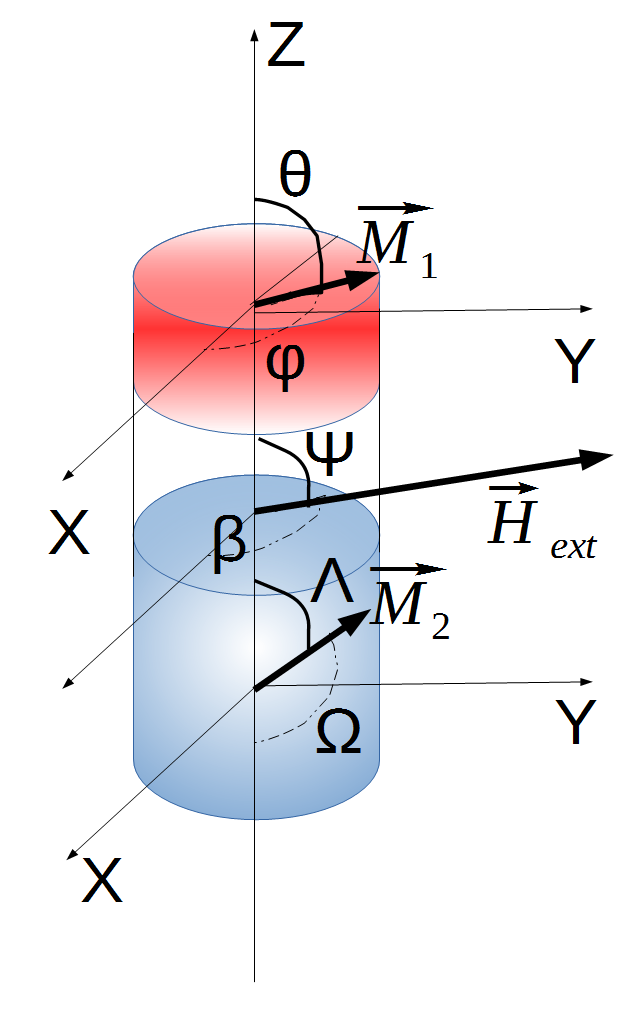}
\caption{Schematics of the magnetic nanowire considered in this paper. Magnetic moments of the top (thin) and bottom (thick) layers are described by the angles ($\theta$, $\phi$) and ($\Lambda$, $\Omega$), respectively. The orientation of the external magnetic field $\vec{H}_{ext}$ (acting on both layers) is described by the angles $\Psi$ and $\beta$.}
\label{fig:geom}
\end{figure}

In this section we present a theoretical description of SDE in the case of two magnetic layers which are dynamically coupled by a dipolar field and STT effects. The system under considerations consists of two magnetic (Co) layers of circular shape, which are separated by a non-magnetic (Cu) layer. The geometry of the nanowire  is shown schematically in Fig.\ref{fig:geom}.

Bearing in mind the SDE, we need to consider first the magnetic dynamics in the system, driven by an {\it AC} charge current flowing along the wire. This dynamics is induced by a time-dependent STT and a time-dependent Oersted field, both being associated with the {\it ac} current. The magnetic moment in each layer is described by the corresponding polar and azimuthal angles,
\begin{equation}
\vec{M}_1 = M_{S,1} [ \sin\theta \cos\phi, \sin \theta \sin\phi , \cos\theta]
\end{equation}
for the thinner layer, and
\begin{equation}
\vec{M}_2 = M_{S,2} [ \sin\Lambda \cos\Omega, \sin \Lambda \sin\Omega , \cos\Lambda]
\end{equation}
for the thicker magnetic layer. Here, $M_{S,1}$ and $M_{S,2}$ are the saturation magnetizations of the thin and thick magnetic layers, respectively.

The magnetization dynamics of the system is described by two coupled Landau-Lifshitz-Gilbert equations,
\begin{subequations}
\begin{align}
\frac{d\vec{M}_1}{dt} = -\gamma_e \vec{M}_1 \times \vec{H}_{\rm eff,1} + \frac{\alpha_1}{M_{S,1}} \vec{M}_1 \times \frac{d\vec{M}_1}{dt} + \gamma_e \vec{\tau_1},\\
\frac{d\vec{M}_2}{dt} = -\gamma_e \vec{M}_2 \times \vec{H}_{\rm eff,2} + \frac{\alpha_2}{M_{S,2}} \vec{M}_2 \times \frac{d\vec{M}_2}{dt} + \gamma_e \vec{\tau_2},
\end{align}
\end{subequations}
where $\gamma_e$ stands for the gyromagnetic ratio, while $\alpha_1$ ($\alpha_2$) and $\vec{\tau_1}$($\vec{\tau_2}$) denote the Gilbert damping parameters and STT vectors acting on the thin(thick) magnetic layer respectively. 
The effective fields $\vec{H}_{\rm eff,1}$ and $\vec{H}_{\rm eff,2}$, in turn,  can be written in the form;
\begin{subequations}
\begin{align}
\vec{H}_{\rm eff,1} = -\nabla_{\theta,\phi} U + H_{\rm oe} \hat{e}_{\phi}, \\
\vec{H}_{\rm eff,2} = -\nabla_{\Lambda,\Omega} U + H_{\rm oe} \hat{e}_{\Omega},
\end{align}
\end{subequations}
where $\hat{e}_{\phi}$ and $\hat{e}_{\Omega}$ are unit vectors associated with the azimuthal angles $\phi$ and $\Omega$, respectively,  $\nabla_{\theta,\phi}$ and $\nabla_{\Lambda,\Omega}$ stand for the relevant gradients in spherical coordinates, $U$ is the total magnetic energy of the system, while $H_{\rm oe}$ is the circular Oersted field generated by the current. Amplitude of the Oersted field, in principle, depends on the distance from the $z$ axis (see Fig.\ref{fig:geom}). However, in the spirit of macrospin model we take into account the strength of the field averaged along the radius of the nanowire.

The total magnetic energy density $U$ can be written as
\begin{eqnarray}
&U(\theta,\phi,\Lambda,\Omega) = K_1 (\cos^2 \theta + \sin^2 \theta \sin^2\phi)  -\vec{M}_1 \cdot \vec{H}_{\rm ext} \nonumber \\
&- \vec{M}_1 \cdot \vec{H}_{dem,1}
 + K_2 (\cos^2 \Lambda + \sin^2 \Lambda \sin^2\Omega) - \vec{M}_2 \cdot \vec{H}_{\rm ext}\nonumber \\
& - \vec{M}_2\cdot \vec{H}_{\rm dem,2}
  - \vec{M}_1 \cdot \vec{H}_{\rm dip,2} - \vec{M}_2 \cdot \vec{H}_{\rm dip,1},\;\;
 \end{eqnarray}
where $K_1$ and $K_2$ are the magnetocrystalline anisotropy constants of the thin and thick layers, respectively,  $\vec{H}_{\rm ext}$ is an external magnetic field,  $\vec{H}_{{\rm dem},1}$ and $\vec{H}_{{\rm dem},2}$ are the demagnetizing fields within the thin and thick layers, while $\vec{H}_{\rm dip,2}$ and
$\vec{H}_{\rm dip,1}$ are the respective dipolar fields. 

Assuming uniform magnetization within both magnetic layers, one can estimate the magnitude of the  interaction between the layers due to dipolar fields by taking into account the field calculated directly at the geometrical center of each layer.
Then, the magnetic moment $\vec{m}$ within volume $dV$ generates the magnetic field that may be expressed as follows:
\begin{equation}
d\vec{H}_{dip} = \frac{1}{\mu_0}\frac{M_S}{4\pi} \left( \frac{3\hat{n} \cdot ( \vec{m} \cdot \hat{n} ) - \vec{m} }{\mathcal{R}^3} \right)
\end{equation}
Here, $\mu_0$ stands for vacuum permeability, $\vec{\mathcal{R}}$  denotes the position vector connecting the volume $dV$ within one magnetic layer and the geometrical center of second layer, while the unit vector $\hat{n}$ is defined as $\hat{n} = \vec{\mathcal{R}}/\mathcal{R}$.
Upon  integrating the above equation over the volume of the whole layer, one finds the dipolar field at any point on the axis $z$:
\newcommand{\Int}{\int\limits}
\begin{equation}
H_{dip,i}(z)= \Int_{0}^{2\pi} d\phi \Int_{0}^{d} dz' \Int_{0}^{R} d\rho'  \rho' dH_{dip,i}
\end{equation}
for  $i \equiv {x,y,z}$, where $d$ and $R$ denote the thickness and radius of the corresponding layer. The above integral may be calculated analytically in our case.  This approach
allows us to calculate dipolar fields in an efficient way without significant overestimations.

Each of the STTs acting on the two magnetic layers, $\vec{\tau}_1$ and $\vec{\tau}_2$, consists of two terms, $\vec{\tau}_1=\vec{\tau}_{\parallel,1}+
\vec{\tau}_{\perp,1}$, and similarly for $\vec{\tau}_2$. These two components can be written as
\begin{subequations}
\begin{align}
\vec{\tau}_{\parallel,1} = \tau_{\parallel,1} ( \hat{m}_{1} \times \hat{m}_{2} \times \hat{m}_{1} ),\label{eq:torki1} \\
\vec{\tau}_{\perp,1} = \tau_{\perp,1} ( \hat{m}_{1} \times \hat{m}_{2} ),\label{eq:torkp1}
\end{align}
\end{subequations}
and
\begin{subequations}
\begin{align}
\vec{\tau}_{\parallel,2} = \tau_{\parallel,2} ( \hat{m}_{2} \times \hat{m}_{1} \times \hat{m}_{2} ),\label{eq:torki2}\\
\vec{\tau}_{\perp,2} = \tau_{\perp,2} ( \hat{m}_{2} \times \hat{m}_{1} ),\label{eq:torkp2}
\end{align}
\end{subequations}
where $\hat{m}_{1}$ and $\hat{m}_{2}$ are unit vectors along the magnetic moments, $\hat{m}_{1}=\vec{M}_1/M_{S,1}$ and $\hat{m}_{2}=\vec{M}_2/M_{S,2}$.
In the spherical coordinate system, the above components  of $\vec{\tau}_1$  may be written as functions of the angles $\theta,\phi,\Lambda,\Omega$ as
\begin{subequations}
\begin{align}
\vec{\tau}_{\parallel,1} = &\tau_{\parallel,1} \left[ \hat{e}_\theta (-\cos\Lambda \sin\theta + \cos\theta\cos(\phi-\Omega)\sin\Lambda)\right. \nonumber \\
&\left.- \hat{e}_\phi \sin\Lambda \sin(\phi-\Omega) \right], \\
\vec{\tau}_{\perp,1} =&\tau_{\perp,1} \left[ \hat{e}_\phi ( \cos\Lambda \sin\theta - \cos\theta \cos(\phi-\Omega) \sin\Lambda ) \right. \nonumber \\
&\left.-\hat{e}_\theta \sin\Lambda \sin(\phi-\Omega) \right],
\end{align}
\end{subequations}
and similarly, the STT components acting on the thick layer may be expressed as
\begin{subequations}
\begin{align}
\vec{\tau}_{\parallel,2} = &\tau_{\parallel,2} \left[ \hat{e}_\Lambda ( \cos\theta\sin\Lambda - \cos\Lambda \cos(\phi-\Omega) \sin\theta ) \right.\nonumber \\
&\left. -\hat{e}_\Omega \sin\theta\sin(\phi-\Omega) \right],\\
\vec{\tau}_{\perp,2} = &\tau_{\perp,2} \left[ \hat{e}_\Omega (  \cos\theta \sin\Lambda - \cos\Lambda \cos(\phi-\Omega) \sin\theta )
\right. \nonumber \\
&\left. -\hat{e}_\Lambda\sin\theta\sin(\phi-\Omega) \right],
\end{align}
\end{subequations}
where $\hat{e}_{\theta}$, $\hat{e}_{\Lambda}$, $\hat{e}_{\phi}$, $\hat{e}_{\Omega}$ are unit vectors associated with the polar  $\theta$, $\Lambda$ and azimuthal $\phi$, $\Omega$ angles, respectively. 

The first components (Eqs.~\ref{eq:torki1} and \ref{eq:torki2}) have the form of damping/antidamping torque and are also frequently described as the in-plane components, because the torque is in the plane determined by the two magnetic moments. The second components (Eqs.~\ref{eq:torkp1} and \ref{eq:torkp2}) are known as the field-like torques as their form is similar to that of the field-generated torque, i.e. the torque is perpendicular to the plane determined by $m_1$ and $m_2$. In metallic systems the field-like torque is usually smaller than the damping/antidamping one.
Apart from this, the scalar amplitudes $\tau_\parallel$ and $\tau_\perp$ may be generally angle dependent due to spin accumulation in the GMR structures.~\cite{boulle2007shaped} In our case, the STT plays the role of an {\it ac} driving force and thus a more complex angular dependence of STT does not lead to significant modifications of the spin diode lineshape or resonance frequency.

The Landau-Lifshitz-Gilbert (LLG)  equations written in spherical coordinates lead to the following equations for the azimuthal and polar angles:
\begin{widetext}
\begin{equation}
\label{eq:llgspher}
\left (
\begin{array}{c}
\dot{\theta}\\
\\
\dot{\phi}\\
\\
\dot{\Lambda}\\
\\
\dot{\Omega}
\end{array}
\right )
=
\left(
\begin{array}{c}
\frac{\gamma}{(1+\alpha_1^2) M_{S,1}}  \left\{ \frac{1}{\sin\theta} \frac{\partial U}{\partial \phi} - \alpha_1 \frac{\partial U}{\partial \theta} -[\cos\Lambda\sin\theta - \cos\theta\sin\Lambda\cos(\phi-\Omega)](\tau_\parallel+\alpha_1\tau_\perp)- \right. \\

\left.- [\sin\Lambda\sin(\phi-\Omega)](\alpha_1\tau_\parallel - \tau_\perp) + M_{S,1} H_{oe} \right\} \\
\\
\frac{\gamma}{(1+\alpha_1^2) M_{S,1}} \left\{ -\frac{1}{\sin\theta}\frac{\partial U}{\partial \theta} -\frac{\alpha_1}{\sin^2\theta}\frac{\partial U}{\partial \phi}+\frac{1}{\sin\theta} [\cos\Lambda\sin\theta - \cos\theta\sin\Lambda\cos(\phi-\Omega)](\alpha_1\tau_\parallel - \tau_\perp) \right. \\
\left. -\frac{1}{\sin\theta} [ \sin\Lambda \sin(\phi-\Omega) ](\tau_\parallel +\alpha_1\tau_\perp) - \frac{\alpha_1 M_{S,1} H_{Oe}}{\sin\theta} \right\}\\
\\
\frac{\gamma}{(1+\alpha_2^2) M_{S,2}}  \left\{ \frac{1}{\sin\Lambda} \frac{\partial U}{\partial \Omega} - \alpha_2 \frac{\partial U}{\partial \Lambda} - [-\sin\Lambda\cos\theta + \sin\theta\cos\Lambda\cos(\phi-\Omega)](\tau_{\parallel,2}+\alpha_2\tau_{\perp,2}) + \right. \\

\left. +[\sin\Lambda\sin(\phi-\Omega)]( \tau_{\perp,2}-\alpha_2\tau_{\parallel,2}) + M_{S,2} H_{oe} \right\} \\
\\
\frac{\gamma}{(1+\alpha_2^2) M_{S,2}} \left\{ -\frac{1}{\sin\Lambda}\frac{\partial U}{\partial \Lambda} -\frac{\alpha_2}{\sin^2\Lambda}\frac{\partial U}{\partial \Omega}+\frac{1}{\sin\Lambda} [\sin\Lambda\cos\theta - \sin\theta\cos\Lambda\cos(\phi-\Omega)]( \tau_{\perp,2} - \alpha_2 \tau_{\parallel,2}) - \right. \\
\left. -\frac{1}{\sin\Lambda} [ \sin\theta \sin(\phi-\Omega) ](\tau_{\parallel,2} - \alpha_2 \tau_{\perp,2}) - \frac{\alpha_2 M_{S,2} H_{Oe}}{\sin\Lambda} \right\}

\end{array}
\right).
\end{equation}
\end{widetext}
Since the induced oscillations of $\vec{M}_1$ and $\vec{M}_2$ around the equilibrium positions can be assumed as small ones, the above equations can be linearized and then their solutions can be expressed as harmonic oscillations,
\begin{subequations}
\begin{align}
\label{eq:rozwa}
&\theta(t)=\theta_0 + \delta\theta (t) = \theta_0 + \delta\theta e^{i(\omega t+\Psi_1)},\\
& \phi(t)=\phi_0 + \delta\phi (t) = \phi_0 + \delta\phi e^{i(\omega t+\Psi_2)} \\
& \Lambda(t)=\Lambda_0 + \delta\Lambda (t) = \Lambda_0 + \delta\Lambda e^{i(\omega t + \Psi_3)} \\
&\Omega(t)=\Omega_0 + \delta\Omega (t) = \Omega_0 + \delta\Omega e^{i(\omega t+\Psi_4)},
\label{eq:rozw}
\end{align}
\end{subequations}
where $(\theta_0, \phi_0)$ and ($\Lambda_0,\Omega_0$) describe the equilibrium orientation of the magnetization in  the thin and thick  layers (in the absence of the {\it ac} driving current),
whereas ($\Psi_{1},\Psi_2$) and ($\Psi_{3},\Psi_4$) take into account the phase shifts in the two magnetic layers between the {\it ac} current and driving force.
The small deviations $(\delta\theta, \delta\phi)$ and ($\delta \Lambda, \delta\Omega$) of the angles from the corresponding equilibrium values are generally complex and include the phase shift between the magnetic dynamics and the driving force.

 After linearization of the RHS of Eq.(\ref{eq:llgspher})  with respect to small changes of the {\it ac} voltage, $\delta V(t)$, and small deviations of $\theta,\phi,\Lambda,\Omega$ from the corresponding stationary  values, one can write  Eq.\ref{eq:llgspher} in the form
 \begin{equation}
 \left (
\begin{array}{c}
\frac{M_{S,1} (1+\alpha_1^2)}{\gamma}\,{\dot{\theta}}\\
\\
\frac{M_{S,1} (1+\alpha_1^2)}{\gamma}\,{\dot{\phi}}\\
\\
\frac{M_{S,2} (1+\alpha_2^2)}{\gamma}\, {\dot{\Lambda}}\\
\\
\frac{M_{S,2} (1+\alpha_2^2)}{\gamma}\, {\dot{\Omega}}
\end{array}
\right )
=
\hat{X}\; \left(
\begin{array}{c}
 \delta\theta(t) \\
 \delta \phi(t) \\
 \delta \Lambda(t) \\
 \delta \Omega(t)
 \end{array}
  \right)
+ \delta V(t)\;\hat{Y},
\label{eq:matrixllg}
 \end{equation}
 where $\hat{X}$ is the $4 \times 4$ matrix consisting of the derivatives of RHS of Eq.\ref{eq:llgspher} with respect to the angles $\theta,\phi,\Lambda,\Omega$, while $\hat{Y}$ is a column matrix composed of the  derivatives of RHS of Eq.\ref{eq:llgspher} with respect to the voltage $V$.

 In the absence of an applied voltage, the second term on the RHS of Eq.(14) vanishes, and the solutions have the form of damped oscillations. In order to calculate resonance frequencies, one can rewrite Eq.\ref{eq:matrixllg} as an eigenvalue problem of the matrix $\hat{A}$, i.e.,
 \begin{equation}
  | \hat{A} - \lambda \hat{I} | = 0,
  \label{eq:czesto}
 \end{equation}
where $\hat{A} = \hat{Z}^{-1} \cdot \hat{X}$ and
\begin{equation}
\hat{Z}=
\left(
\begin{array}{cccc}
\frac{M_{S,1} (1+\alpha_1^2)}{\gamma} & 0 & 0 & 0 \\
0 & \frac{M_{S,1} (1+\alpha_1^2)}{\gamma} & 0 & 0 \\
0 & 0 & \frac{M_{S,2} (1+\alpha_2^2)}{\gamma} & 0 \\
0 & 0 & 0 & \frac{M_{S,2} (1+\alpha_2^2)}{\gamma}
\end{array}
\right).
\nonumber
\end{equation}
The eigenvalues $\lambda$ determine the resonance frequencies of the system, $\omega_i = \Im \lambda $.
In general, a $4\times 4$ matrix has four eigenvalues, however we will show that in our system only two different eigenfrequencies occur.

Now we find from Eq.(\ref{eq:matrixllg}) a general solution for $\delta\theta$, $\delta\phi$, $\delta\Lambda$ and $\delta\Omega$ in the presence of applied {\it ac} voltage.
To do this we rewrite  Eq.\ref{eq:matrixllg} in matrix form as
  \begin{equation}
 \hat{B}\, \hat{\Psi}\, \hat{\Gamma} e^{i\omega t} = \hat{Y}\, \delta V e^{i\omega t},
 \label{eq:vdc2}
  \end{equation}
 where $\hat{B}=\hat{Z}-\hat{X}$, $\hat{\Psi}\equiv \mathrm{diag} (e^{i\Psi_1},e^{i\Psi_2},e^{i\Psi_3},e^{i\Psi_4})$  and $\hat{\Gamma}\equiv (\delta\theta,\delta\phi,\delta\Lambda,\delta\Omega)^T$.
 After dividing both sides of the above equation by the oscillating term $e^{i\omega t}$, one obtains a set of linear equations  for  $\delta\theta, \delta\phi, \delta\Lambda, \delta\Omega$  (or for the matrix $\hat{\Gamma}$), from which one finds  $\hat{\Gamma}$ in the form
\begin{equation}
\hat{\Gamma} = \left( \hat{B}\, \hat{\Psi} \right)^{-1} \, \hat{Y} \, \delta V .
\end{equation}
Finally, we replace $\hat{\Gamma}$ by its real part $\Re \{ \hat{\Gamma} \}$.

Having found the general formulas describing magnetization dynamics, we can now determine the spin diode voltage $V_{DC}$  and the lineshape of the spin diode signal. First, we recall that the resistance of a GMR spin valve depends on the angle $\xi$ between magnetizations $\vec{M}_1$ and $\vec{M}_2$ as $R(\xi) = R_P + \Delta R(1-\cos\xi)/2$, where $R_P$ is the resistance in the parallel configuration, while $\Delta R$ is the difference in resistance of antiparallel and parallel configurations.
Thus, the spin-diode signal for the GMR nanowire can be expressed by a small change $\delta \xi$ of the angle $\xi$, which is induced by magnetic dynamics. Accordingly, we write the spin-diode signal as:~\cite{ziketek2015rectification}
\begin{equation}
V_{DC} = \frac{1}{2} I\; \delta R,
\label{eq:vdc1}
\end{equation}
where $I$ is the amplitude of {\it ac} current, while $\delta R$ is the amplitude of resistance change,
\begin{equation}
\delta R(\xi) = \frac{\Delta R}{2} \sin\xi \delta \xi .
\end{equation}

The angle $\xi$ can be considered as  a function of the angles $\theta$, $\phi$, $\Lambda$, and $\Omega$.
Due to magnetization dynamics, the angles $\theta,\phi,\Lambda,\Omega$
change by $\delta\theta$, $\delta\phi$, $\delta\Lambda$ and $\delta\Omega$, respectively. Thus, the change of $\xi$  due to small changes of the angles
$\theta,\phi,\Lambda,\Omega$ can be written as
\begin{equation}
\delta\xi = \frac{\partial \xi}{\partial \theta}\delta\theta+\frac{\partial \xi}{\partial \phi}\delta\phi
+\frac{\partial \xi}{\partial \Lambda}\delta\Lambda+\frac{\partial \xi}{\partial \Omega}\delta\Omega
\label{eq:ksi}
\end{equation}
The above equation, together with the solution (17) for the angle changes $\delta\theta$, $\delta\phi$, $\delta\Lambda$, and $\delta\Omega$, is sufficient to find the spin diode signal.

\section{Experimental\label{sec:experiment}}

To perform measurements of the $V_{DC}$ signal in a GMR system with two magnetically free layers, we prepared asymmetric spin valves consisting of two layers of cobalt, one thicker and one thinner, separated by a copper layer. A standard electrodeposition process in commercial ion-track etched polycarbonate membranes was used. The membrane was 5 $\mathrm{\mu m}$ thick and contained nanopores of 30 nm in diameter with a pore density of 6 $\times$ 10$\mathrm{^6/cm^2}$. Details of the methods for growing and contacting the nanowires were similar to those described in Refs \onlinecite{biziere2009microwave,mure2010lithography,fitoussi2015linear}. The membrane was placed on top of the central conductor of an SMA connector. A shorting plug was mounted on the connector, that contained a screw at the tip of wich a gold wire was soldered. The wire would rub the surface of the electrodeposited membrane until contact was established. An optical fiber was driven through a hole on the side of the connector and was pointing at the membrane. This allowed us to drive the ferromagnetic resonance with microwave currents at 4 to 10 GHz. The rectified signal (through the spin diode effect) was detected by a standard lock-in technique, using the drive of the field modulation coil or of the laser diode as reference.

Equation~(\ref{eq:vdc1}) for the $V_{DC}$ signal has to be modified due to the measurement technique used in the experiment. This technique is  commonly known as the field-modulated FMR.~\cite{gonccalves2013spin} Furthermore, we explored here the possibility to apply temperature modulation, as a simple laser diode is sufficiently powerful to modulate the temperature of a nanowire.~\cite{gravier2004thermopower}

 The $V_{DC}$ signal in both techniques can be expressed simply as:
 \begin{equation}
V_{DC,H}= \frac{d V_{DC}}{d H_m} \delta H_m
 \label{eq:fieldfmr}
  \end{equation}
 and
 \begin{equation}
 V_{DC,T} = \frac{d V_{DC}}{d T} \delta T ,
 \label{eq:tempfmr}
  \end{equation}
 where $\delta H_m$ and $\delta T$ denote the amplitude of modulation field and the temperature gradient, respectively.
 In this model, $V_{DC}$ signal does not depend on the temperature explicitly. However, it is possible to introduce temperature-dependent anisotropy and magnetization according to Bloch's law.
 In this case, we may express Eq.\ref{eq:tempfmr} as:
$$V_{DC,T} = \frac{d V_{DC}}{d T} \delta T = $$
\begin{equation}
= \sum_{i=1}^2 \left( \frac{\partial V_{DC}}{\partial M_{S,i}}\frac{\partial M_{S,i}}{\partial T}+
\frac{ \partial V_{DC}}{\partial K_i}\frac{\partial K_i}{\partial T} \right)  \delta T .
  \label{eq:tempfmr2}
  \end{equation}
  In the following $V_{DC,H}$ and $V_{DC,T}$ will be denoted simply as $V_{DC}$, if not stated otherwise.

\section{Results\label{sec:results}}


\begin{figure*}
\includegraphics[width=18cm]{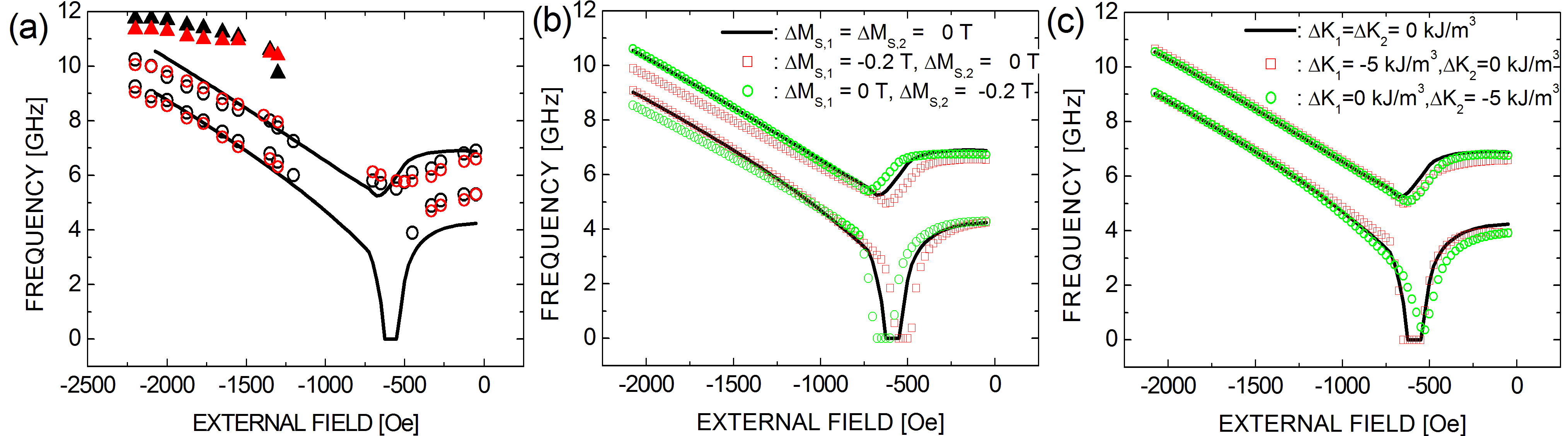}
\caption{The resonance frequencies versus external field: (a) the theoretical curves (solid lines)  calculated from Eq.(\ref{eq:czesto}), compared to experimental points in the field modulation (black points) and laser modulation (red points) technique, (b) theoretical resonance frequency shift due to a temperature dependence of the saturation magnetizations as well as (c) due to temperature dependence of the magnetocrystalline anisotropy.}
\label{fig:dysp}
\end{figure*}
\begin{figure*}
\includegraphics[width=18cm]{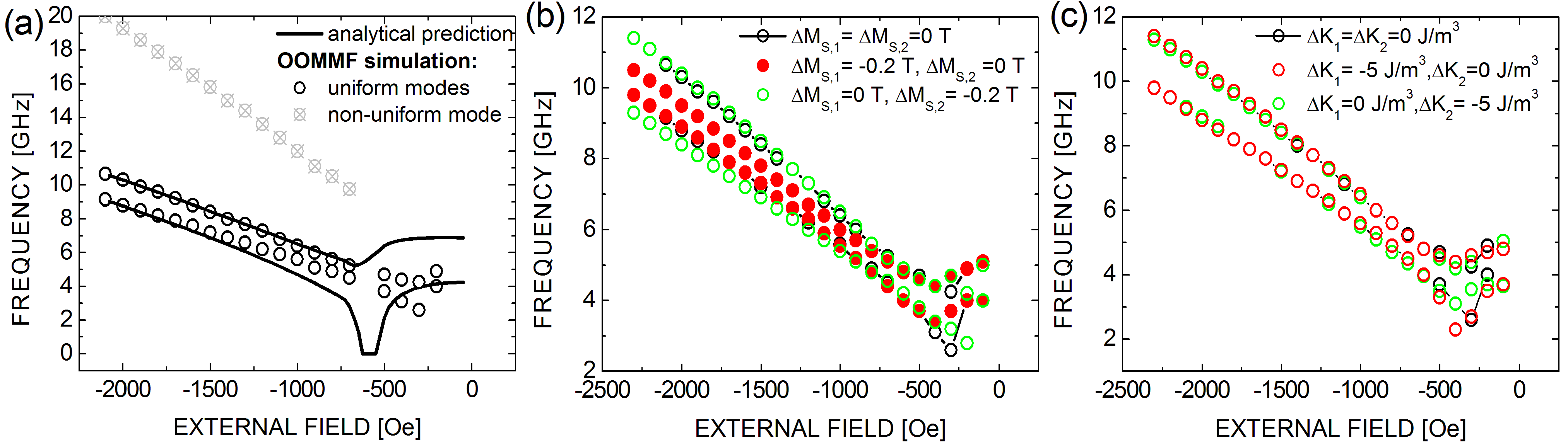}
\caption{Micromagnetic simulations: (a) resonance modes versus external field  (circles) qualitatively compared to the macrospin predictions (black solid lines), (b) resonance frequencies shifts due to temperature-induced decrease of magnetization and (c) due to temperature-induced decrease of  magnetocrystalline anisotropy. }
\label{fig:dysp-sym}
\end{figure*}

The nanowires under consideration consist of two cobalt ferromagnetic layers separated by a non-magnetic Cu spacer. Because of the electrochemical deposition method used to fabricate the nanowires, the thin layer in an individual nanowire may have slightly different magnetic properties than the corresponding thick layer. This is due to the presence of copper atoms that are co-deposited in the magnetic layers. As a consequence, the thinner layer can be expected to be more affected by the non-magnetic impurities. For our purposes, we assume the following magnetic parameters of both layers:
$M_{S,2} =  1.4 T, K_2 = 25 kJ/m^3$ and $M_{S,1}=1.2 T, K_1 = 15 kJ/m^3$.
Apart from this, the thicknesses of thick and thin layers were chosen as $d_2=12 nm$ and  $d_1 = 5 nm$, respectively. In turn, for the thickness of the non-magnetic Cu spacer, we assumed $25$ $nm$, for which the mean values of dipolar fields  in the thick ($\bar{H}_{dip,1}$) and thin ($\bar{H}_{dip,2}$) layers were 168 Oe and 488 Oe, respectively.

In the experiment, the nanowires are deposited into the nanopores of a polycarbonate mambrane. Thus, the external magnetic field applied in the plane of the whole membrane may not be applied exactly in the plane of nanowire's cross section. In other words, the nanowire axis
may not be collinear with the membrane axis, and they may form an angle of up to 40 degrees. Similarly, the external field orientation may not be exactly perpendicular to the easy axes of the magnetic layers, but it may deviate from 90 degrees. The polar and azimuthal angles describing the orientation of  the external magnetic field, $\Psi$ and $\beta$, have been set as $\Psi = 72^\circ$ and $\beta = 63^\circ$ to ensure a good agreement with the experimentally measured  dispersion relation as well as to obtain a non-zero angle between the magnetic moments even at high external magnetic fields.





\subsection{Dispersion relation}

In Fig.\ref{fig:dysp}(a) we compare the theoretical dispersion relation for the modes determined from Eq.\ref{eq:czesto} with the experimental measurements done with both FMR detection schemes: field modulation and temperature modulation. In general, there are two ways in which laser heating can affect the sample. One of them is a laser-induced decrease in the magnetocrystalline anisotropy, and the second one is a laser-induced decrease in the magnetization according to the Bloch's law applied to nanostructures.~\cite{cojocaru2014}
 As one can see in Fig.\ref{fig:dysp}(b) and (c), the influence of the increase in temperature on resonance frequencies is rather small and only a slight shift of the frequency can be seen. Interestingly, a similar slight shift can be observed also in the experimentally determined dispersion relations  as well as in those obtained from the micromagnetic simulations shown in Figs. \ref{fig:dysp}(a) and \ref{fig:dysp-sym}(b)-(c). The simulations were performed with the use of OOMMF package\citep{donahue1999oommf} with our custom software.\citep{frankowski2014micromagnetic,chkecinski2016mage} The LLG equation was solved numerically in each 2$\mathrm{\times}$2$\mathrm{\times}$1 nm simulation cell for magnetic parameters identical to those utilized in the analytical calculations described above. After the initial magnetization relaxation, a small magnetic pulse of 10 Oe was used to excite both ferromagnetic layers and their magnetization response was subjected to a Fast Fourier Transform (FFT) from which we obtained the resonance frequencies of the system.

The analytical model based on the macrospin approximation predicts only two separate resonance modes. However, the experimental observations and micromagnetic simulations of the frequency {\it vs} magnetic field dependence indicate the presence of a third mode at high magnetic fields, see Fig.\ref{fig:dysp}(a) and Fig.\ref{fig:dysp-sym}(a). This additional mode may be related to a more complex dynamics in the system -- possibly due to presence of magnetic inhomogeneities at the boundaries of both magnetic layers -- or to standing-like spin-wave excitations.  However, since we focus in this paper on the influence of different modulation techniques on the SDE lineshapes of the FMR modes, the non-uniform mode remains beyond the scope of this paper. Thus, in the following we will discuss the FMR (uniform) modes only, for which the experimental, macrospin and micromagnetic approaches are consistent, cf. Fig.\ref{fig:dysp} and Fig.\ref{fig:dysp-sym}.

\subsection{Lineshapes in both field and laser modulation techniques}
\begin{figure}[H]
\includegraphics[width=8.5cm]{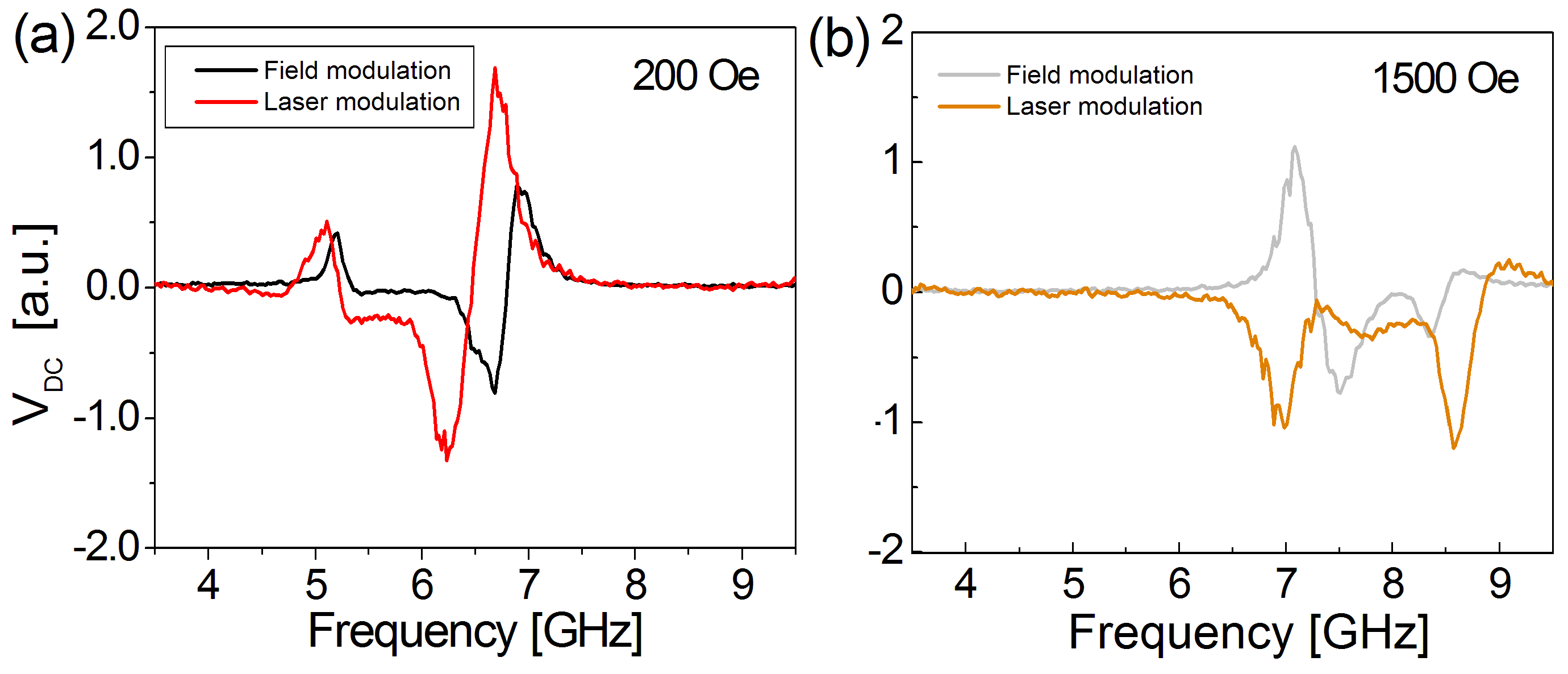}
\caption{Experimental lineshapes measured at low field 200 Oe (a) and high field 1.5 kOe (b) with the two modulation techniques: magnetic field modulation (black and light grey lines) and laser modulation (red and orange lines).}
\label{fig:linia0}
\end{figure}

The experimentally observed  $V_{DC}$ spectra are presented in Fig.\ref{fig:linia0}. As one can note, the symmetry of the peaks in the field modulation and laser modulation techniques is the same at low fields. On the contrary, the symmetry of the peaks at high fields is different in the two modulation techniques. This observation raises the question as to why the modulation technique affects the lineshape symmetry. This problem is now analyzed in detail.

The lineshape of the $V_{DC}$ spectrum is very sensitive to the phase shifts and to the origin of the driving force. Due to strong coupling of the layers by both the dipolar field and STT effects, the symmetries of both peaks cannot be considered separately. Instead, we look at the simultaneous dynamics of both ferromagnetic layers, and consider the symmetry of the whole spectrum. For example, a change of the phase shifts for the magnetization dynamics in one layer has a significant influence on the lineshape of the two resonant peaks, as they are linked to each layers. 

\begin{figure}[H]
\includegraphics[width=8.5cm]{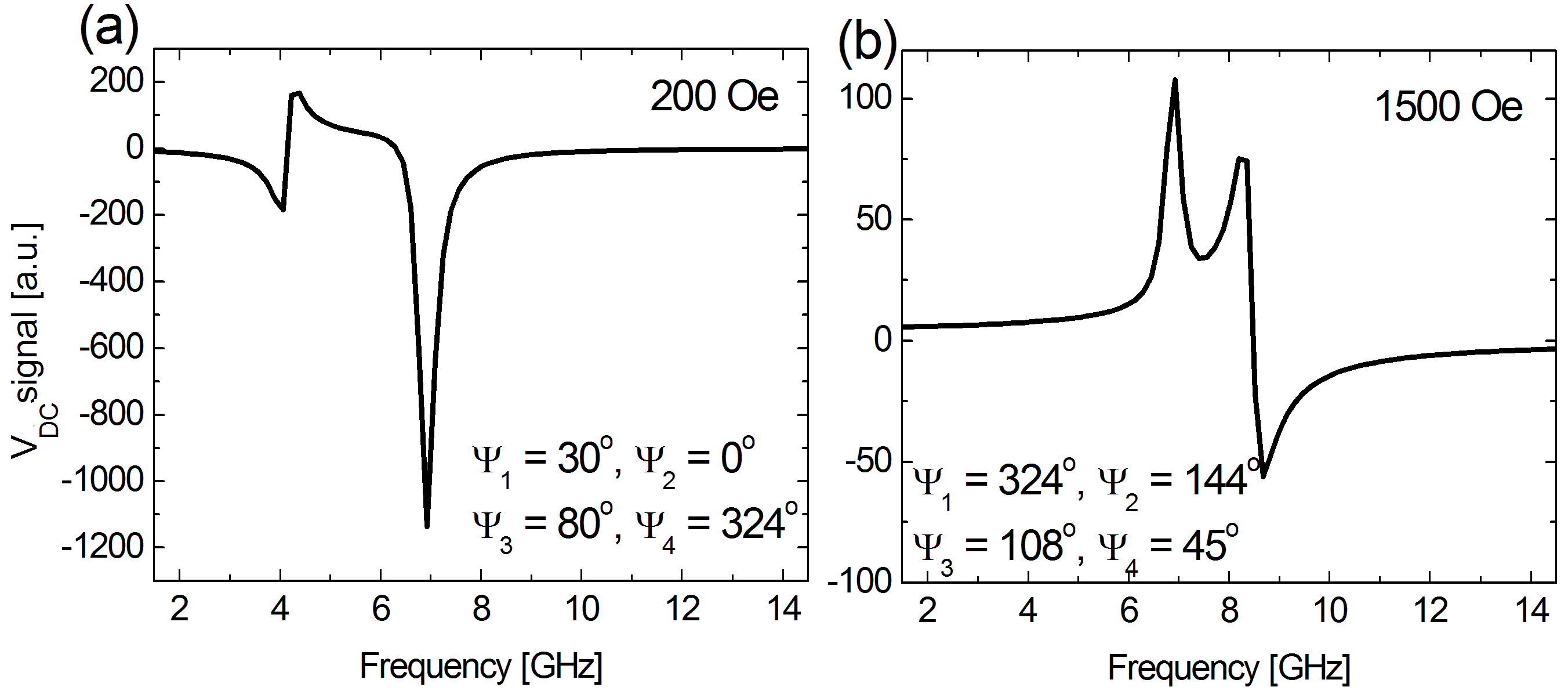}
\caption{The $V_{DC}$ lineshape in the mode with no modulation,  calculated at external field 200 Oe (a) and 1500 Oe (b). The amplitudes of STT in both layes and the amplitude of small Oersted field were assumed arbitrary.}
\label{fig:linia1}
\end{figure}

\begin{figure}[H]
\includegraphics[width=8.5cm]{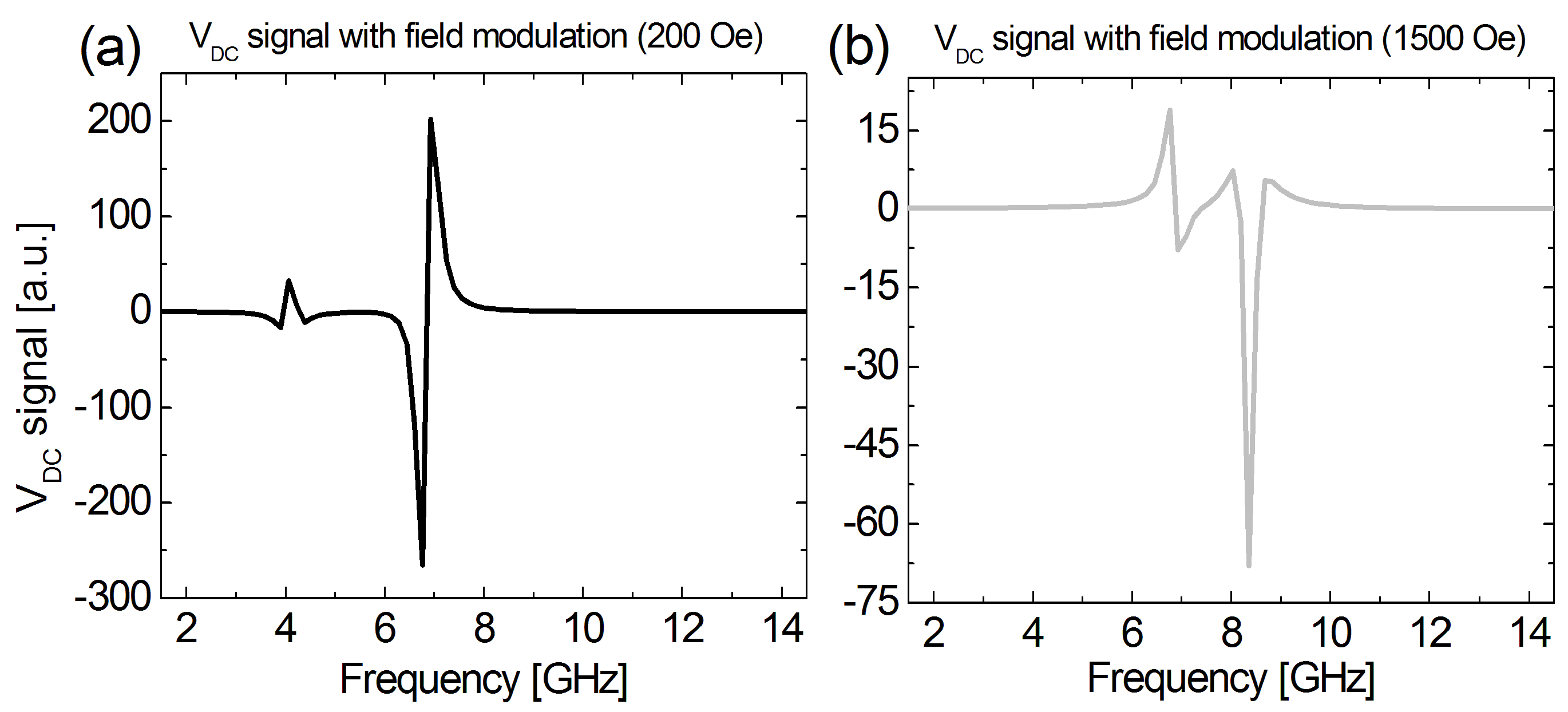}
\caption{The $V_{DC}$ lineshape in mode with field modulation calculated at external field 200 Oe (a) and 1500 Oe (b). All the parameters are the same as in Fig.\ref{fig:linia1}.}
\label{fig:linia2}
\end{figure}

Let us start the analysis of lineshapes  with choosing the phase shifts $\Psi_{1,2,3,4}$ to obtain agreement between the theoretical lineshapes and the experimental ones in the field modulation technique.  According to Eq.\ref{eq:fieldfmr}, the output $V_{DC}$ voltage is a derivative with respect to the modulating magnetic field. Thus a symmetric (antisymmetric) peak which is detected with the field modulation is antisymmetric (symmetric) when no field modulation is applied. In Fig.\ref{fig:linia1} we plot the unmodulated $V_{DC}$ signal at low and high external magnetic fields. The phase shifts indicated  in Fig.\ref{fig:linia1}(a) and (b) are used throughout this paper, so we do not discuss their influence on the spectra, but we consider only the influence of the modulation technique. The signal with field modulation is shown in Fig.\ref{fig:linia2}. As one can conclude from this figure, the lineshapes agree well with the experimental ones at both low and high fields.

The other modulation technique used is the laser modulation one. In this method, the rectified signal $V_{DC}$ is determined as a derivative with respect to the amplitude of the temperature modulation (induced by laser). According to Eq.(\ref{eq:tempfmr2}), in order to calculate properly the $V_{DC}$ signal in the laser modulation technique one has to know the temperature dependence of $M_{S,i}$ and $K_{i}$.  Here, we omit this problem and assume that the saturation magnetization and anisotropies decrease with increasing temperature, i.e. their derivatives are negative. We do not consider the magnitude of the derivatives since our objective is to show how the change of modulation technique influences lineshapes. For the sake of simplicity we assume that the changes in the magnetic parameters (saturation magnetization and anisotropy constants) in both layers are the same and that the phase shifts are the same as those assumed in the case of the field modulation technique.


\begin{figure}[H]
\includegraphics[width=8.5cm]{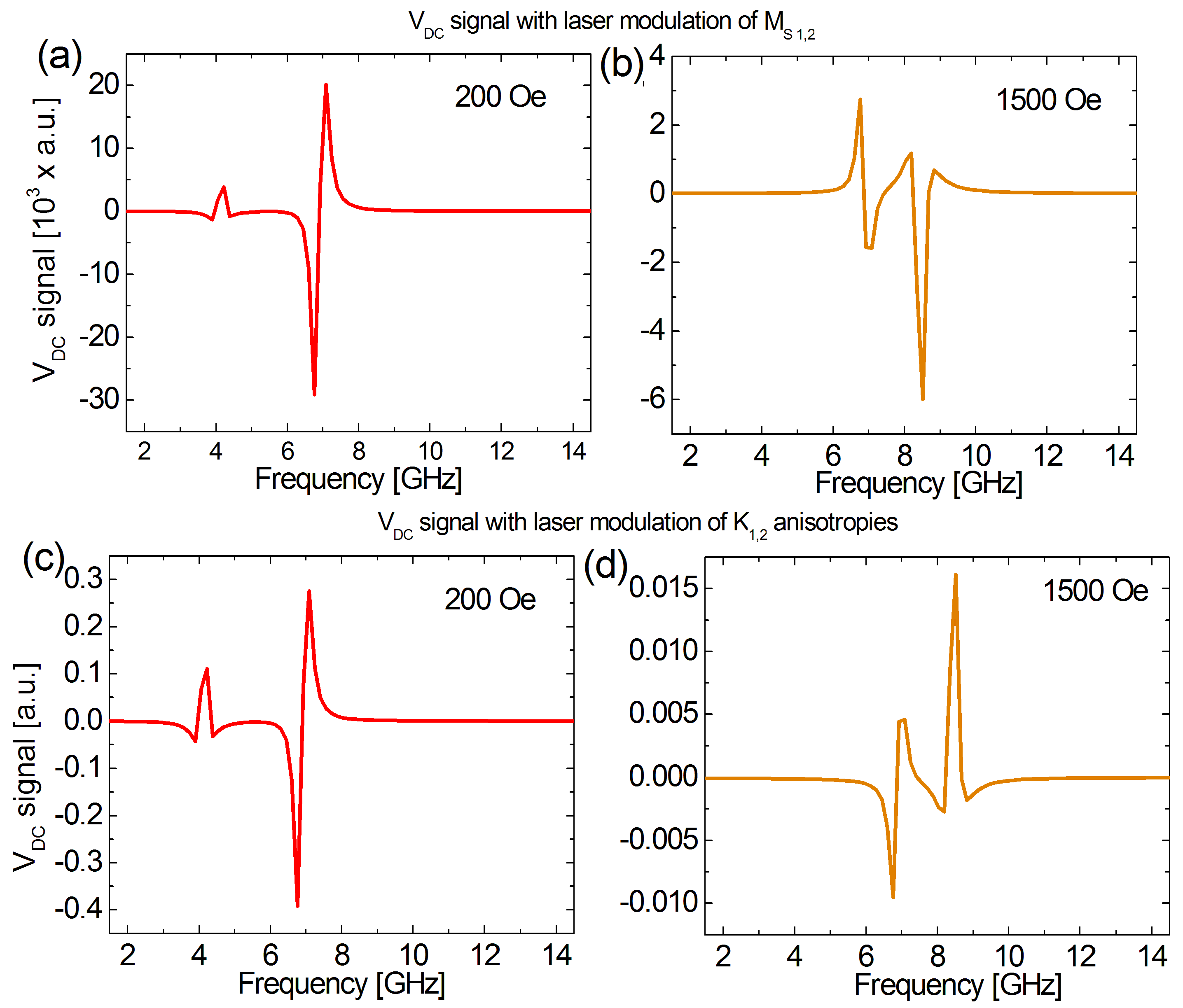}
\caption{The $V_{DC}$ lineshape with laser (temperature) modulation of the magnetization (a,b) and anisotropies (c,d), calculated at external field 200 Oe and 1500 Oe. The phase shift and other parameters are the same as in the field modulation technique}
\label{fig:linia3}
\end{figure}

In Fig.\ref{fig:linia3} we show the $V_{DC}$ signal in the laser modulation technique. We consider two cases: the first case is the modulation of saturation magnetizations, while the second one is the modulation of anisotropies. As one can see, in the first case there is  no  change of the lineshape, but the amplitude of the signal is much higher. However, one should bear in mind that the amplitude depends on the derivative $\frac{\partial M_{S,i}}{\partial T}$ and modulation temperature $\delta T$ as well. Thus, the real amplitude of the signal will be smaller than that shown in this figure. On the contrary, the temperature modulation of the anisotropies affects the lineshape (see Fig. \ref{fig:linia3}(c) and (d)). Furthermore, in contrast to the field modulation (Fig.\ref{fig:linia2}(a)), the first peak is higher with respect to the second one. A similar enhancement of the first peak at low fields is also present in the experimental data.

In Fig.\ref{fig:linia3}(b) and (d) we show the $V_{DC}$ signal for temperature-induced modulation of $M_{S,i}$ and $K_i$ at high magnetic field. The lineshape is slightly different in the case of $M_{S,i}$ modulation mode than in the field modulation mode. On the contrary, the $V_{DC}$ signal changes the sign when  $K_{i}$ is modulated at high field. Thus one can say that at low field the symmetry of the lineshape is conserved in both field and temperature modulation. Furthermore, the lineshape is more affected by temperature-induced anisotropy modulation than by saturation magnetization modulation at high fields.

\begin{figure}[H]
\includegraphics[width=8.5cm]{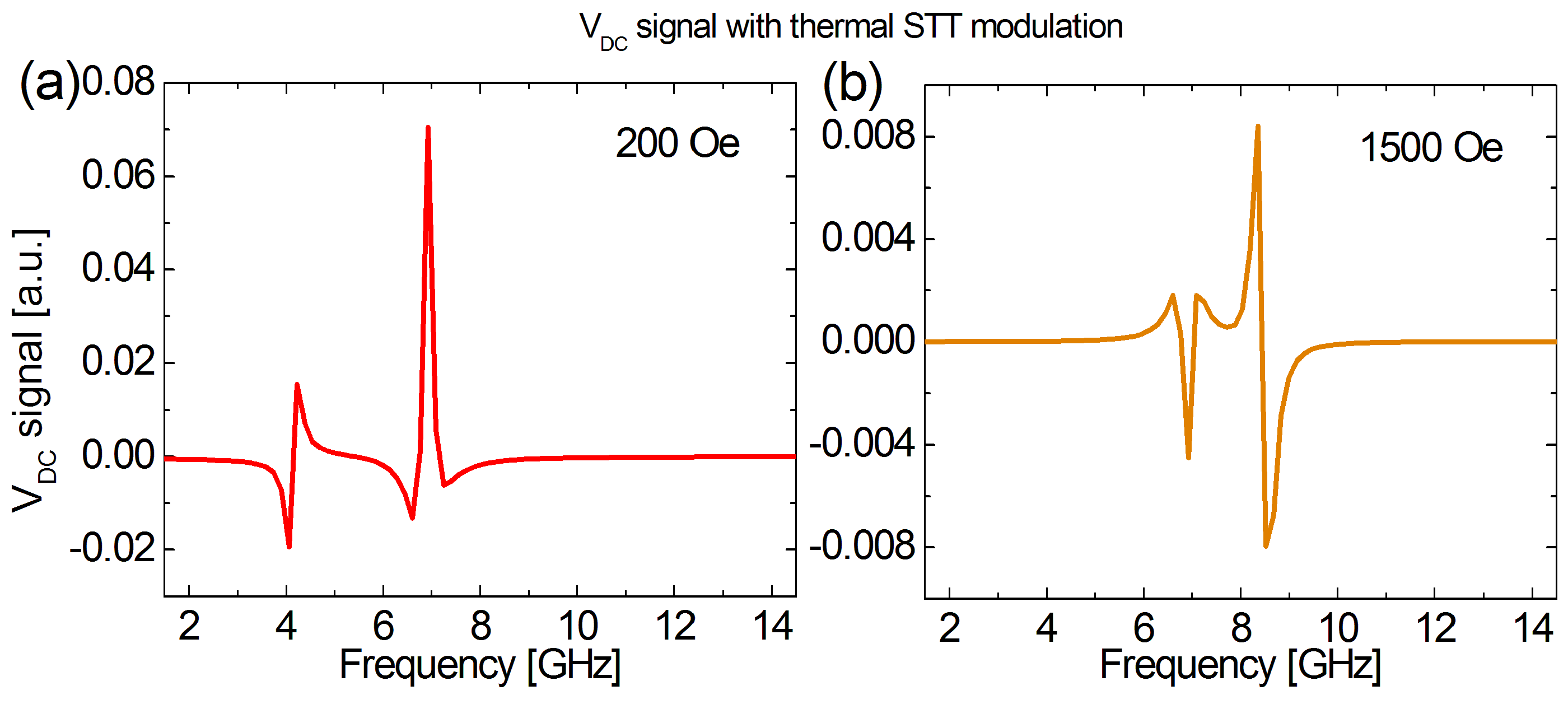}
\caption{The $V_{DC}$ lineshape with laser (temperature) modulation of thermal STT calculated at low field (200 Oe) (a) and high field (1500 Oe)  (b). The phase shifts and other parameters are the same as in the field modulation mode.}
\label{fig:linia5}
\end{figure}

We now turn to the question whether the laser-induced temperature gradient gives rise to a significant heat-driven spin torque (thermal STT). We assume (for simplicity) that a small thermal in-plane torque has the same amplitude in both layers. The rectified signal with thermal STT modulation may be expressed as follows:

\begin{equation}
V_{DC,T} = \sum_i \frac{\partial V_{DC}}{\partial \tau_{\parallel,i} }\frac{\partial \tau_{\parallel,i}}{\partial T} \delta T
 \label{eq:tempfmrtork}
\end{equation}
In Fig.\ref{fig:linia5} we can see how such a type of modulation influences the lineshape of the $V_{DC}$ signal.

As one can see, the thermal STT modulation changes the symmetry of both peaks: symmetric (antisymmetric) peaks in field modulation become antisymmetric (symmetric) peaks in the thermal STT modulation. These results differ from those observed experimentally. There may be several reasons for this. One of them are phase shifts, that in a general case do not have to be the same in the field and laser modulation techniques. Second, the amplitude of thermal STTs also may differ in the two layers. Third, the laser heating may influence only one layer, e.g. the free layer. This is justified since  the 3/2 Bloch-like law for nanostructrures states that $M_{S,i}$ is size- and shape-dependent\cite{cojocaru2014}, and thus the thick and thin layers may be characterized by  different derivatives, i.e.: $\frac{\partial M_{S,1}}{\partial T} \neq \frac{\partial M_{S,2}}{\partial T}$. A similar inequality may appear for $K_1$ and $K_2$ temperature dependences. For example, the derivatives related to one of the magnetic layer may vanish, and then only one of the layers is affected by temperature. This seems to be important in the case of laser modulation of saturation magnetization and anisotropy. Finally, all of the above reasons may be present and may compete with each other.

\begin{figure}[H]
\includegraphics[width=8.5cm]{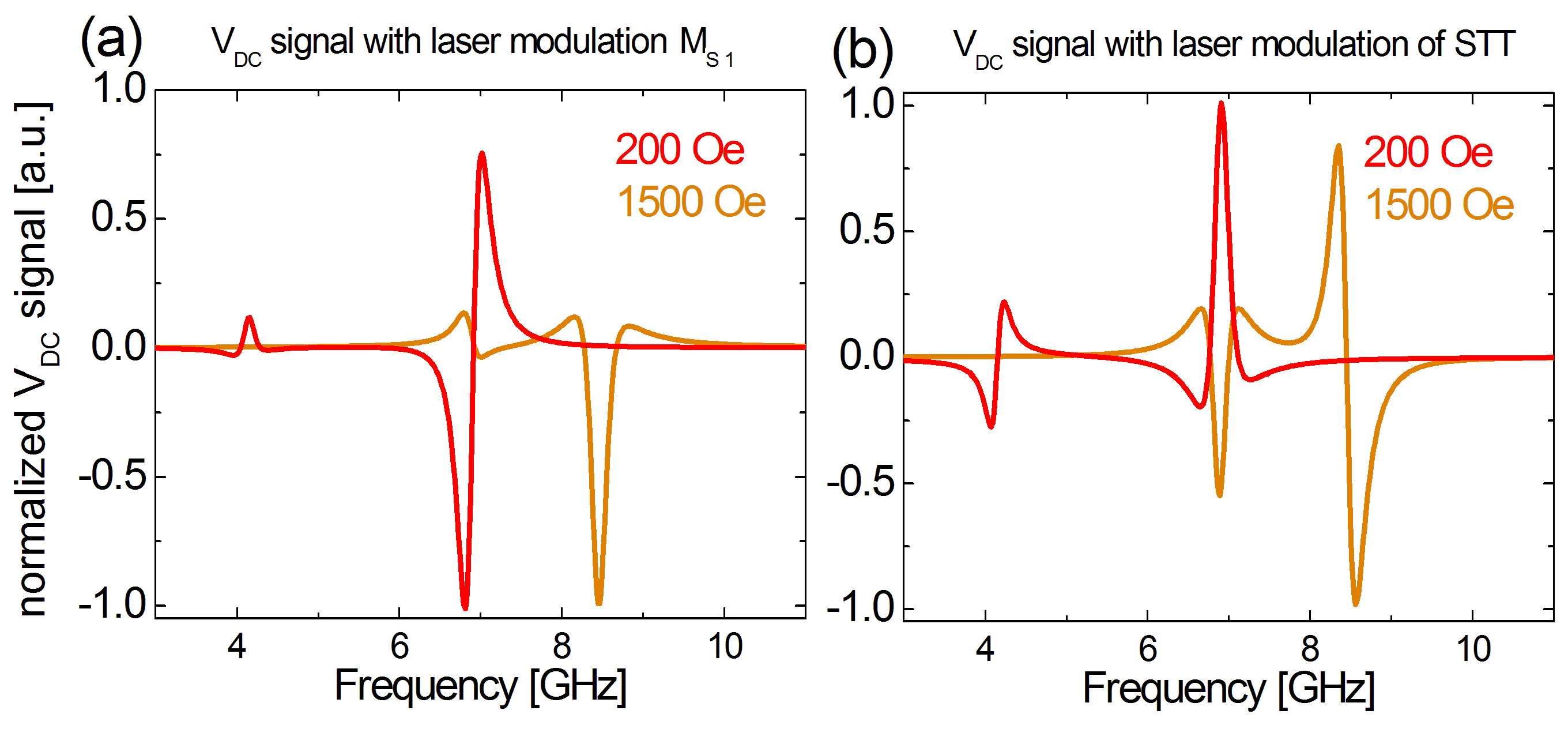}
\caption{The normalized $V_{DC}$ lineshape with laser (temperature) modulation of magnetization of the free layer ($\vec{M}_1$) (a) and thermal STT (b) calculated at low field (200 Oe) and high fields. The phase shifts and other parameters are the same as in the field modulation mode.}
\label{fig:linia6}
\end{figure}
\begin{figure}[H]
\includegraphics[width=8.5cm]{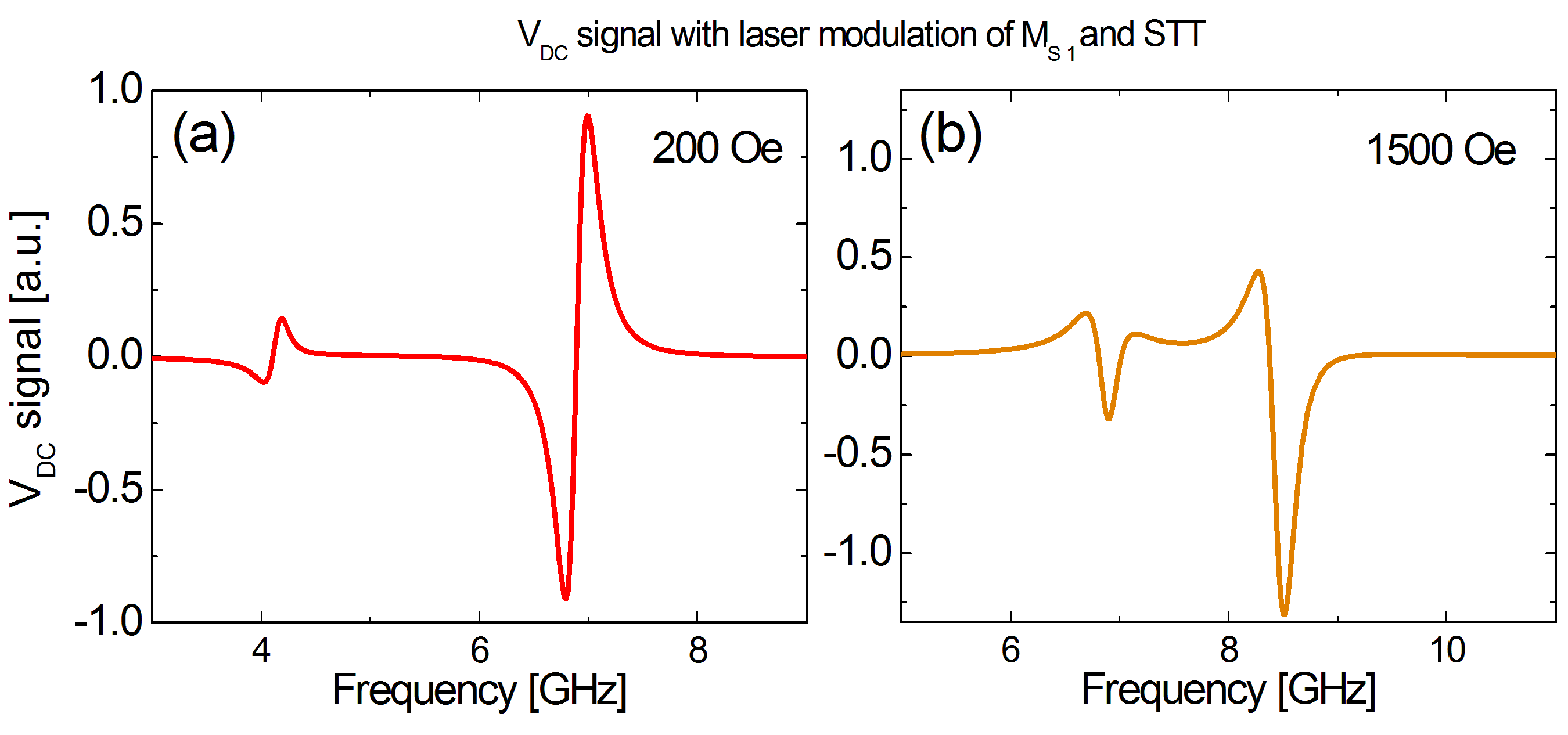}
\caption{$V_{DC}$ lineshape with contribution from laser (temperature) modulation of magnetization of the free layer ($\vec{M}_1$)  and thermal STT calculated at (a) low field (200 Oe) and (b) high fields (1500 Oe). The phase shifts and other parameters are the same as in the field modulation mode.}
\label{fig:final}
\end{figure}

To illustrate these issues, we calculated lineshapes in the case when the magnetization of the free layer is modulated only. The normalized $V_{DC}$ lineshapes are shown in Fig.\ref{fig:linia6}(a). The ratio of the amplitude of second peak to the amplitude of first peak at low field is almost the same as in Fig.\ref{fig:linia3}(a). However, this ratio is much higher at high field (cf. Figs.\ref{fig:linia3}(b) and \ref{fig:linia6}(a)). Next, we normalized the result from Fig.\ref{fig:linia5} and depicted it in Fig.\ref{fig:linia6}(b). In order to obtain more realistic spectra, we need to consider thermal modulations (of $M_{S,1}$ and STT) simultaneously. We can do it by adding spectra from Fig.\ref{fig:linia6}(a) and (b) with different weights (in our case weights are 1.0 and 0.35 respectively).

The result is shown in Fig.\ref{fig:final}. As one can see in Fig.\ref{fig:final}(a), at low field the lineshapes are still similar to those measured in experiment. At high field (Fig.\ref{fig:final}(b)), both peaks are more symmetric than when the modulations of ${M}_{S,1}$ and STT are considered separately. This means that during the laser heating all kinds of thermal modulations may contribute, leading effectively to different lineshapes.

\section{Conclusions\label{sec:concl}}

In this paper we discussed the important factors pertaining to the SDE measurements in GMR nanowires with two magnetic layers coupled by both dipolar field and STT effects. First, we showed that our simple analytical model is able to describe experimentally observed FMR (uniform) modes in this system. Our results were supported by micromagnetic simulations as well. Next, we calculated the SDE lineshapes in two different, field and laser (heat), modulations techniques. We compared our analytical predictions with experimental SDE spectra. We found that experimental lineshapes can be successfully analysed by the present model. Moreover,  we discussed the influence of the modulation method on resonance peaks symmetries, and showed that the modulation technique may importantly affect the SDE lineshapes. We considered different contributions to the SDE spectra coming from heat-related modulation of magnetizations, anisotropies and STTs. We showed that the temperature-modulated magnetic anisotropy influences the SDE lineshapes more than the temperature-modulated magnetization. However, a possible competition of different heat-related phenomena makes the lineshape analysis in laser modulation technique much more complex than in the case of field modulation technique. Our model turned out to be helpful in recognizing which physical quantity is mostly modulated during laser heating.  We believe that our contribution  may be useful for analysis of further SDE experiments, especially for those combining spintronic with caloritronic effects. The presented model may be used to investigate spin-diode spectra in different GMR or TMR structures, where dynamics of two strongly coupled layers has to be taken into account.

\section{Acknowledgments\label{sec:acknow}}
The research was financed by NANOSPIN Grant No. PSPB-045/2010 from Switzerland
through the Swiss Contribution. Numerical calculations were supported in part by PL-GRID infrastructure. J.Ch. acknowledges the scholarship under Marian Smoluchowski Krakow Research Consortium KNOW programme. T.S. acknowledges National Science Center (NCN), Poland (Grant No.2011/02/A/ST3/00150). P.O. acknowledges Dekaban Fund at the University of Michigan for the financial support.

\bibliographystyle{apsrev4-1}

\bibliography{bibliography}

\end{document}